\documentclass[sigconf]{acmart}

\usepackage{graphicx}
\usepackage{multirow}
\usepackage{subfigure}
\usepackage{soul}
\usepackage{longtable}

\AtBeginDocument{%
  \providecommand\BibTeX{{%
    \normalfont B\kern-0.5em{\scshape i\kern-0.25em b}\kern-0.8em\TeX}}}

\setcopyright{rightsretained}
\copyrightyear{2021}
\acmYear{2021}
\acmDOI{10.1145/3461778.3462086}

\acmConference[DIS 2021]{ACM DIS Conference on Designing Interactive Systems}{June 28-- July 2, 2021}{Virtual event, USA}
\acmBooktitle{ACM DIS Conference on Designing Interactive Systems,
  June 28-- July 2, 2021, Virtual event, USA}
\acmPrice{?}
\acmISBN{978-1-4503-8476-6/21/0}

\begin{document}

\title[Aware but manipulated]{"I am Definitely Manipulated, Even When I am Aware of it. It’s Ridiculous!" - Dark Patterns from the End-User Perspective}

\author{Kerstin Bongard-Blanchy}
\email{kerstin.bongard-blanchy@uni.lu}
\orcid{0000-0001-9139-1622}
\affiliation{%
  \institution{University of Luxembourg}
  \streetaddress{2, avenue de l'Université}
  \city{Esch sur Alzette}
  \country{Luxembourg}
  \postcode{4366}
}
\author{Arianna Rossi}
\email{arianna.rossi@uni.lu}
\orcid{0000-0002-4199-5898}
\affiliation{%
  \institution{SnT, University of Luxembourg}
  \streetaddress{29, Avenue J. F. Kennedy}
  \city{Luxembourg}
  \country{Luxembourg}
  \postcode{1855}
}
\author{Salvador Rivas}
\email{salvador.rivas@uni.lu}
\orcid{0000-0002-1496-6004}
\affiliation{%
  \institution{University of Luxembourg}
  \streetaddress{2, avenue de l'Université}
  \city{Esch sur Alzette}
  \country{Luxembourg}
  \postcode{4366}
}
\author{Sophie Doublet}
\email{sophie.doublet@uni.lu}
\affiliation{%
  \institution{University of Luxembourg}
  \streetaddress{2, avenue de l'Université}
  \city{Esch sur Alzette}
  \country{Luxembourg}
  \postcode{4366}
}
\author{Vincent Koenig}
\email{vincent.koenig@uni.lu}
\orcid{0000-0001-9940-6286}
\affiliation{%
  \institution{University of Luxembourg}
  \streetaddress{2, avenue de l'Université}
  \city{Esch sur Alzette}
  \country{Luxembourg}
  \postcode{4366}
}
\author{Gabriele Lenzini}
\email{gabriele.lenzini@uni.lu}
\orcid{0000-0001-8229-3270}
\affiliation{%
  \institution{SnT, University of Luxembourg}
  \streetaddress{29, Avenue J. F. Kennedy}
  \city{Luxembourg}
  \country{Luxembourg}
  \postcode{1855}
}

\renewcommand{\shortauthors}{Bongard-Blanchy, et al.}

\begin{abstract}
Online services pervasively employ manipulative designs (i.e., dark patterns) to influence users to purchase goods and subscriptions, spend more time on-site, or mindlessly accept the harvesting of their personal data. To protect users from the lure of such designs, we asked: are users aware of the presence of dark patterns? If so, are they able to resist them? By surveying 406 individuals, we found that they are generally aware of the influence that manipulative designs can exert on their online behaviour. However, being aware does not equip users with the ability to oppose such influence. We further find that respondents, especially younger ones, often recognise the "darkness" of certain designs, but remain unsure of the actual harm they may suffer. Finally, we discuss a set of interventions (e.g., bright patterns, design frictions, training games, applications to expedite legal enforcement) in the light of our findings.
\end{abstract}

\ccsdesc[500]{Security and privacy~Social aspects of security and privacy}
\ccsdesc[300]{Security and privacy~Usability in security and privacy}
\ccsdesc[500]{Human-centered computing~Empirical studies in HCI}
\ccsdesc[100]{Human-centered computing~Graphical user interfaces}

\keywords{dark patterns, online manipulation, digital nudging, consumer protection, user experience, user interface}

\maketitle

\section{Introduction}
The pervasiveness of manipulative practices in online services is increasingly under the limelight. Thanks to information technologies, manipulative practices can be implemented at low costs, at large scale, with unprecedented sophistication \cite{Mathur2021what} and high effectiveness \cite{Susser2018manipulation} in dynamic, interactive, intrusive, and adaptive environments \cite{Susser2019technology}. Such online practices endeavour to influence purchase decisions, nudge people to spend considerable amounts of time on a service (thus intensifying data collection to fuel the so-called attention economy \cite{attention_economy}), and trick users into accepting privacy-invasive features, thereby undermining their right to the protection of their personal information and exposing them to privacy harms. Online manipulation does not only erode legal protections, but it also deprives unaware individuals of their capacity for independent decision-making \cite{Susser2019technology}. Therefore, the phenomenon is scrutinised by a growing number of practitioners and researchers, with the aim of exposing it and devising countermeasures. 

This article focuses on a specific form of online manipulation: dark patterns. They are defined as ``design choices that benefit an online service by coercing, steering or deceiving users into making decisions that, if fully informed and capable of selecting alternatives, they would not make'' \cite{mathur2019dark}. Dark patterns direct user behaviour towards choices that may offer advantages like ease of use, free services and immediate gratification. However, such choices may have an adverse impact on individual welfare (e.g., invasion of privacy, financial loss, behavioural addiction) and collective welfare (e.g., harm to competition, erosion of consumer trust) \cite{Mathur2021what}. Dark patterns are believed to work because they exploit cognitive biases and human bounded rationality \cite{bosch2016tales,waldman2020cognitive}, and a growing body of research demonstrates the influence of dark patterns on online behaviours. Although experts voice their apprehension, users' awareness regarding dark patterns is still an under-researched topic and has only been recently the object of dedicated studies \cite{Maier2020dark,gray2018dark}.  

The study presented in this article seeks to fill this gap, by determining whether dark patterns exploit (1) users' lack of awareness or concern; (2) users' incapability of recognising dark patterns (the so-called ``dark-pattern blindness'' \cite{di2020ui}); or (3) users' inability to resist dark patterns, despite their awareness and ability to recognise them. By investigating the user perspective, we aim to identify requirements for effective countermeasures, that can either act on the individual or on factors lying outside the individual. If users are unaware or unconcerned about dark patterns, one solution consists in strengthening their motivation to counteract them (e.g., using warnings to increase the salience of risks). Suppose users are concerned about the risks deriving from dark patterns, but they are nevertheless unable to withstand them. In that case, their ability to resist needs to be improved (e.g., adding friction designs that disrupt automatic behaviour), while stronger environmental protections should also be leveraged (e.g., steep fines against companies employing dark patterns).

This article makes the following contributions to understand users' awareness of manipulative designs online: (1) It reveals that users are able to recognise dark patterns, but they are only vaguely aware of the entailed concrete harm. It furthermore hints that a higher ability to discern manipulative designs is positively related the capacity to self-protect. Moreover, the study shows that people under 40 and with higher education than high school diplomas are more likely to recognise dark patterns.
(2) The findings guide designers, educators, developers, and regulators to draft appropriate interventions to counteract manipulative designs online, both in terms of intervention scope and measure.
\label{sec:intro}

\section{Related work}
Previous work has investigated the presence of dark patterns in online services  \cite{matte2020cookie,soe2020circumvention,human2020human,di2020ui,Conti2010malicious}, even through automated means like web scraping \cite{mathur2019dark,nouwens2020dark,kampanos2021landscape}. Various categories and definitions have been proposed to characterise the phenomenon in general \cite{web_dp1,gray2018dark}, but also specifically to account for video-games \cite{Zagal2013games}, ubiquitous computing \cite{greenberg2014dark}, automated systems \cite{michael2019dark} and home robots \cite{Lacey2019cuteness}; or in the context of data privacy \cite{bosch2016tales,chatellier_shaping_2019} and e-commerce \cite{mathur2019dark}. Examples of dark pattern implementations have been gathered in online collections \cite{web_dpgames,web_confirmsh,web_dp1,web_dp2,web_dppolitics} to create knowledge, raise awareness, propose alternatives and build training corpora for algorithms. Although there is a consensus that dark patterns can employ and even combine coercive, deceptive, and nudging strategies, clear boundaries between (inadmissible) manipulative designs and other (admissible) designs (e.g., digital nudges helping users reach praiseworthy goals, like adopting more secure behaviour online) are yet to be set. \citet{Mathur2021what} recently sought to bring coherence to the existing jungle of definitions and attributes by establishing that some dark patterns modify the decision space in an asymmetric, restrictive, unequal or covert manner. In contrast, others manipulate the information flow through deception or the concealment of information. They also map out the normative considerations that underpin the problematic nature of dark patterns with respect to other designs. They diminish the individual and the collective welfare, weaken regulatory objectives and undermine individual autonomy. 

A growing number of studies demonstrates the effect of dark patterns on online behaviour and strives to find the causes of their effectiveness, especially of those extorting consent in cookie dialogues \cite{utz2019informed,nouwens2020dark,grassl2020dark,machuletz2020multiple,soe2020circumvention,human2020human}. It has been proposed \cite{bosch2016tales,waldman2020cognitive} that human innate cognitive limitations (e.g., cognitive biases, bounded rationality) are skilfully exploited by online services to direct users toward choices they may regret \cite{Stigler2019}. Examples are the status quo bias that benefits from the human tendency to stick with the default option and the bandwagon effect that leverages herd behaviour. Certain cognitive biases might interfere with risk assessment \cite{barth2017privacy} and can hereby explain ill-decision making. For example, hyperbolic discounting causes people to overvalue current rewards (e.g., accomplish a task), while they inadequately discount the cost of future risks \cite{waldman2020cognitive} (e.g., privacy invasion). The optimism bias \cite{sharot2011optimism} might make individuals underestimate their disposition to online manipulation. 

Some scholars investigated whether individuals are able to identify dark patterns. \citet{di2020ui} introduced the notion of "dark pattern-blindness", to explain why most respondents (i.e., well-educated, of various origins) in their study were not able to recognise dark patterns in mobile applications. However, when the study participants were informed of the potential presence of dark patterns in the context at hand, they became more capable of spotting them. \citet{luguri2019shining} showed that mild (i.e., more subtle) dark patterns go more easily unnoticed than aggressive ones and that less-educated individuals are significantly more likely to be influenced than more educated subjects. \citet{Shaw2019wise_nudge} noticed how the overuse of scarcity and social proof messages on travel websites makes consumers ignore them even on other types of websites. On a similar note, \citet{Bhoot2020IndiaCHI} found that the ability to identify a dark pattern is correlated with its frequency of occurrence and the frustration it provokes. If the interface is appealing, respondents tend to experience less frustration and hardly notice manipulative attempts. The experimental data shows that certain design attributes can influence people's capacity of spotting and resisting dark patterns.

The attitudes of various stakeholders towards dark patterns have been explored, too. Design practitioners \cite{web_dp1,falbe2020handbook,Brownlee2016why,Jaiswal2018dark,Singer2016when} have been the first to voice their concerns over these questionable practices. Similar considerations have been developed by regulators \cite{chatellier_shaping_2019,acm2020guidelines} and consumer organisations \cite{forbrukerradet_deceived_2018,Forbruker2021amazon}. Several studies \cite{gray2018dark,chivukula2019analyzing,chivukula2020coevolving,chivukula2020dimensions,watkins2020tensions,gray2019ethical,gray2020kind} have analysed practitioners' ethical values and their conflict with other stakeholders' interests. However, only recently it has been inquired whether dark patterns are a source of concern for end-users. \citet{Maier2020dark} found a raising awareness and a general sense of annoyance among Swedish students. They also uncovered resignation, as their respondents believed it impossible to avoid online manipulation, and they acknowledged that the benefits (e.g., free service) outweigh the negative consequences. In an analogous study \cite{Gray2020enduser}, English-speaking and Mandarin-speaking respondents evoked a general impression of manipulation in digital products. They were able to identify what makes them grow suspicious, even though they lacked a specific vocabulary to indicate the source of that feeling.
 
 In terms of solutions, \citet{grassl2020dark} used design nudges (so-called bright patterns) to reverse the direction of dark patterns and steer users' consent decisions towards the privacy-friendly option (e.g., pre-selection of "Do not agree" option). They also recommended long-term boosts that help users acquire procedural rules because the repeated use of analytic thinking converts into protection heuristics (e.g., every time I encounter a consent request, I take the time to read the information before making a choice). Based on a survey among impulse buyers, \citet{Moser2019impulse} proposed friction designs that counteract dark pattern mechanisms in purchase decisions (e.g., disabling urgency and scarcity messages). \citet{Bhoot2020IndiaCHI} and \citet{mathur2019dark} suggested a plug-in or browser extension that automatically detects dark patterns on websites and notifies the user.  \citet{Leiser2020regulatory} discussed the regulatory tools that can be leveraged to prohibit and fine these practices. In parallel, \citet{Maier2020dark} assumed that dark patterns diminish customers' trust in and the credibility of a brand in the long term, leading customers to stop using the service.\label{sec:relatedwork}

\section{Research gaps and research questions}
The spectrum of possible dark pattern design implementations is vast, ranging from coercive designs that constrain user options to nudges that subtly play on the visual prominence of one choice over another. Thus, it is impossible to identify one single intervention that could free the web from all dark patterns. Drafting appropriate interventions is hence a design problem in itself. Before working on the solutions, it is however indispensable to understand which user issue the interventions aim to solve. 

Individuals may execute a threat appraisal \cite{Rogers1997protection} that makes them believe that dark patterns do not inflict serious harm. They may also think that they are invulnerable - or at least less vulnerable than others, as it is customary in online risk appraisal \cite{west2008psychology}. We therefore asked:
\begin{itemize}
    \item[RQ1] Are users \textit{aware of} and \textit{concerned about} the influence of manipulative interface designs on their behaviour?
\end{itemize}  
 
 \citet{di2020ui} concluded that individuals are subject to 'dark pattern-blindness'. It is also assumed that manipulation is a hidden influence, while coercion is not \cite{Susser2018manipulation}, and that nudges work only when people are unaware of the influence that is exerted on them \cite{lembcke2019towards}. This is why we asked:  \begin{itemize}
     \item[RQ2] Are users \textit{able to recognise} manipulative interface designs?
 \end{itemize}
 
 It is further assumed that the transparency (i.e., the visibility) of an influence is a crucial dimension for its acceptability, because it gives the opportunity to control the influence \cite{hansen2013nudge}, for instance by resisting. However, transparency may not be sufficient to contrast the influence: resignation, benefits (e.g., free service) \cite{Maier2020dark}, the cognitive costs of opposing dark patterns and other factors might undermine the ability to resist. This is why we sought to relate awareness, ability to detect and influence of dark patterns on users, by asking:
\begin{itemize}
    \item[RQ3] Are users likely to be \textit{influenced} by manipulative interface designs despite being aware of, concerned about, and capable of recognising manipulative interface designs?
\end{itemize}
Additionally, lower  educational levels seem to be correlated with a greater influence on consumer behaviour \cite{luguri2019shining}. Therefore, we explored if level of education, age, and use frequency of online services are significantly associated with the three research questions.\label{subsec:research_q}

\section{Approach}
\subsection{Study design}
\begin{figure}
    \centering
    \includegraphics[scale=0.1]{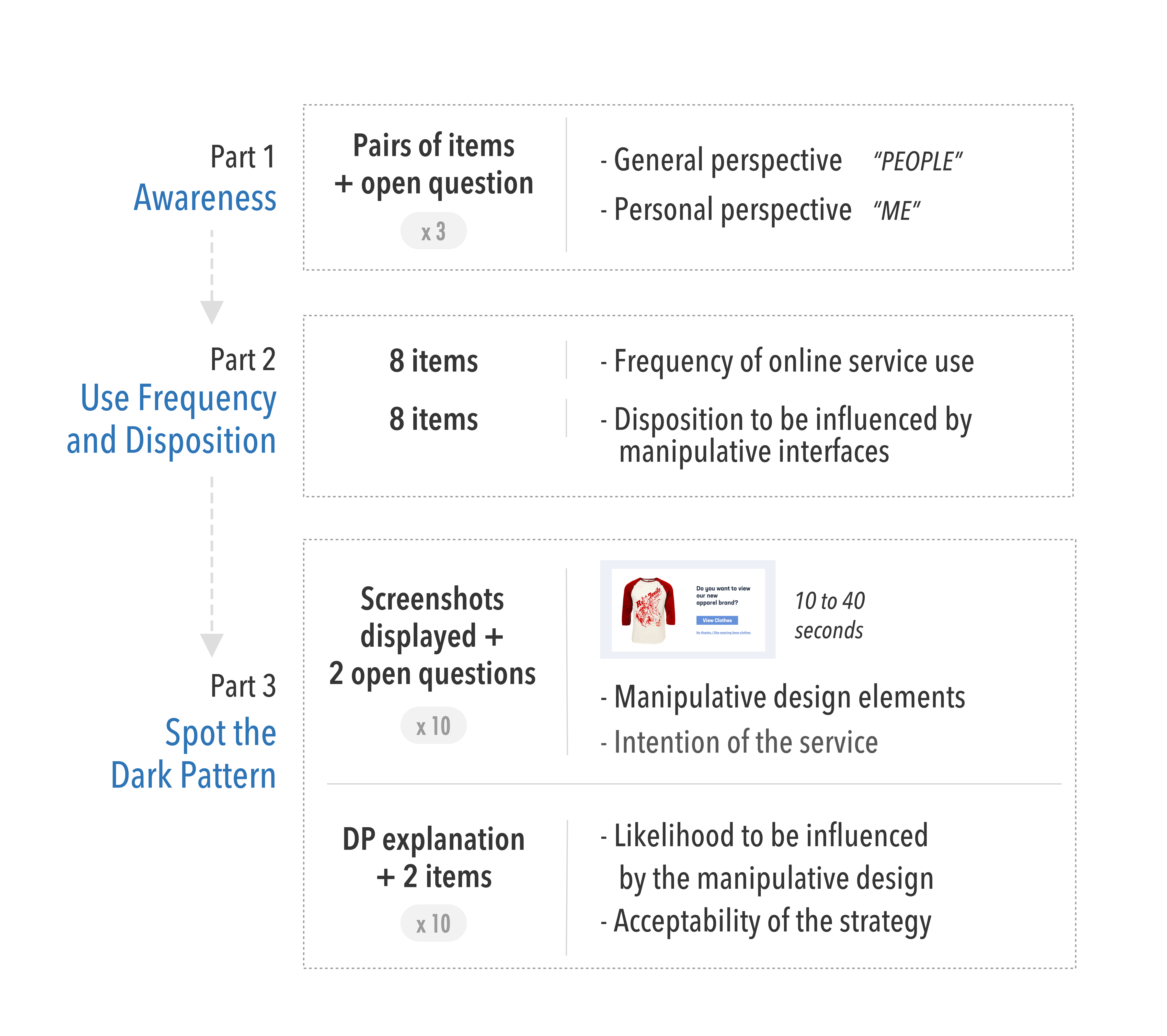}
    \caption{Sequence of the questions in the survey.}
    \label{fig:surveysequence}
\vspace{-3mm}
\end{figure}

To investigate the three research questions, we designed an online survey on LimeSurvey,\footnote{\url{https://www.limesurvey.org/}} administered through Prolific.\footnote{\url{https://www.prolific.co/}} People were first asked about their general mindset concerning manipulative designs online, followed by ratings of their online behaviour, before being exposed to specific dark pattern designs (Figure \ref{fig:surveysequence}). In addition, demographic data regarding their gender, age, and education was gathered. All questions were mandatory, except for a final general feedback field. The three parts of the survey are detailed in the following.

\subsubsection{Part 1: Awareness and concern}\label{meth-Part1:awareness}
The first part of the survey addressed participants' awareness and concerns about online designs' potential influence. Six statements were displayed in pairs (one pair per page) opposing general perspective (``people/others'') and personal perspective (``my/me''):

\begin{itemize}
\item The design of websites or applications can influence [people's/my] choices and behaviours.
\item Websites or applications that are designed to manipulate\footnote{In the second and third question of part 1, the term "manipulate/manipulative" was chosen to indicate dark patterns in a commonly understandable manner, while in the first question "influence" was preferred to avoid negative priming. Similarly, we used both terms in Part 3: Spot the dark pattern. In certain cases, we deliberately chose the term manipulation because influence of design can be very widely interpreted and lead to answers that are not relevant for the research at hand (as the answers to the first open question illustrate).} users can cause harm to [people/me].
\item I am worried about the influence of manipulative websites and applications on [people's/my] choices and behaviours.
\end{itemize}

Participants were instructed to rate their agreement on a 5-point Likert scale (from -2 = \textit{strongly disagree} to 2 = \textit{strongly agree}). If they gave an affirmative or undecided answer (0, 1 or 2), they were furthermore invited to cite examples of experienced influence, potential harm and related worries after each statement pair.

\subsubsection{Part 2: Use frequency of online services and disposition to manipulation}\label{meth-Part2:behaviour} The second part served to complement the demographic data. To obtain a proxy of participants' exposure to online services, participants had to rate the frequency of their engagement with eight common services (\textit{How often do you: play online games / order products online / use social media / etc.?}). As a second indicator, we sought to obtain the participants' disposition to be influenced by manipulative designs online. To this end, participants had to indicate their usual behaviour in eight situations in which web services commonly employ manipulative strategies \textit{(While using online services: I reserve a service quickly when there are only a few items left. / I keep the default permissions when I install an app. / etc.)}. We randomised the item order for both tasks and kept the phrasing neutral to avoid that the action would be perceived as undesirable behaviour.

\subsubsection{Part 3: Spot the dark pattern}\label{meth-Part3:spot}
\begin{figure*}
    \centering
    \includegraphics[scale=0.64]{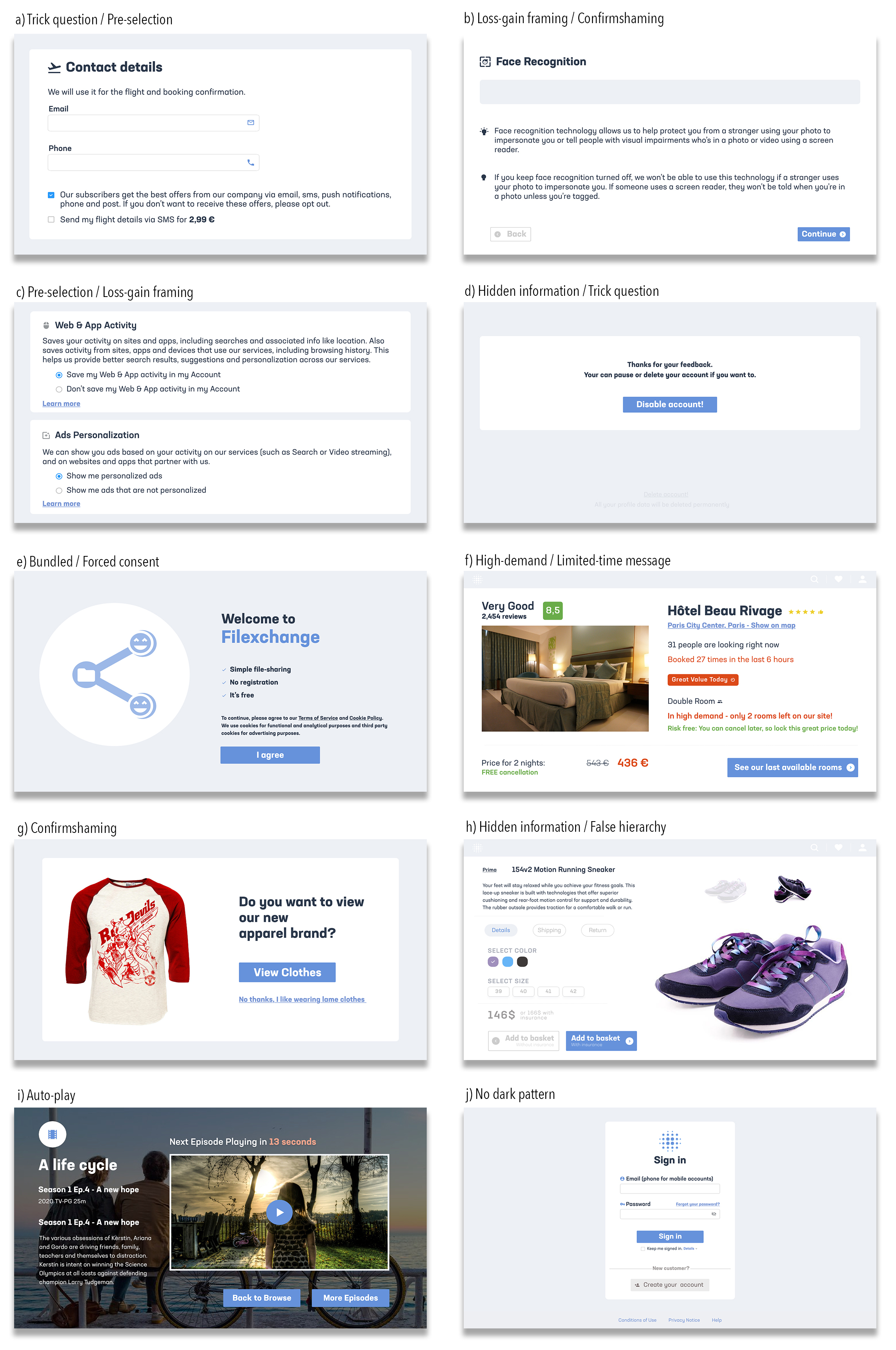}
    \caption{The interfaces designs tested in this study.}
    \label{fig:DPexamples_details}
\end{figure*}

The third part served to evaluate the participants' capability to recognise different dark pattern types. Ten interfaces of existent online services were displayed in random order. The interfaces had been redesigned in a uniform style and freed of any reference to a real brand. One example without any dark pattern was included as control condition. The other nine examples contained dark patterns that impact individual welfare, causing financial harm, data privacy harm, and time and attention-related harm. Within these categories, authors one and two gathered numerous examples of existing interfaces with dark patterns from reports \cite{forbrukerradet_deceived_2018}, online collections\footnote{E.g., \url{https://www.reddit.com/r/darkpatterns/}} and personal screenshots. The interface selection represented a mix of popular and less known brands of various services, such as e-commerce websites, dating apps, and social media. Table \ref{table:DPdef} provides the definition of the dark patterns embedded in the interfaces shown in Fig. \ref{fig:DPexamples_details}.

\begin{table*}[ht]
    \begin{tabular}{rp{11cm}}
     Type & Definition  \\ 
     \hline
     \textbf{High-demand message} & Indicating that a product is in high demand and likely to sell out soon \cite{mathur2019dark}  \\
     \textbf {Limited-time message} & Indicating that a deal will expire soon without specifying a deadline \cite{mathur2019dark} \\
     \textbf {Confirmshaming} & Using shame to steer users towards making a certain choice \cite{mathur2019dark} \\
     \textbf {Trick question} &  Using confusing language to steer users towards making a certain choice \cite{mathur2019dark}  \\
     \textbf {Loss-gain framing} & A selective disclosure of information that positively frames the consequences of an action, while omitting the entailed risks \cite{boush2009deception}  \\
     \textbf {Pre-selection} & An option is selected by default prior to user interaction, even though it is  against her interest or may have unintended consequences \cite{gray2018dark}    \\
     \textbf {False hierarchy} & Visual or interactive prominence of one option over others,  whereas available choices should be evenly leveled rather than hierarchical  \cite{gray2018dark}  \\
     \textbf {Hidden information} & Disguising relevant information (options, actions) as irrelevant \cite{gray2018dark}    \\
     \textbf {Auto-play} & Automatically loading one video when the previous ends \cite{alter2017irresistible}   \\
      \textbf {Bundled consent} & Gathering consent for multiple settings through a single action (\textit{our own definition} but see \cite{santos2020cookie} )  \\
     \textbf {Forced consent} & Coercing users into accepting fixed legal terms in exchange for access to the service \cite{chatellier_shaping_2019}   \\
      \hline
    \end{tabular}
    \caption{Definitions of the dark patterns selected for this study.}
    \label{table:DPdef}
    \vspace{-5mm}
\end{table*}

Each example was displayed for 10 to 40 seconds, depending on its textual complexity. The participants were asked if they noticed any design element that might influence their behaviour. It was made explicit beforehand that not all examples contained such elements. This indication and the time constraint served to limit excessive searching that does not occur in a regular use context. Once the image disappeared, the participants saw a thumbnail of the interface and a text field. They had to describe the manipulative element (i.e., the means of the influence) and the presumable service intention (i.e., its ends) to employ that element. After going through all ten interfaces, the participants were given an explanation about the contained dark pattern(s). The explanations also pointed to potential benefits of these designs for users (e.g., ease of use). To conclude, the participants had to rate on a 5-point Likert scale (from -2 = \textit{strongly disagree} to 2 = \textit{strongly agree}) if they believed it likely to be influenced by the displayed designs and if they considered the strategy employed by the online service acceptable.

\subsection{Participants}
The survey collected responses from 413 participants. The data of seven individuals that gave gibberish answers were excluded from data analysis, leaving a sample size of 406. Prolific allowed to gather a representative sample of the UK population in terms of age, gender and ethnic origin.\footnote{\url{https://researcher-help.prolific.co/hc/en-gb/articles/360019238413}} Since this option was only available for the UK and the US, the former was selected to address participants living in a uniformly regulated digital ecosystem. The demographics of the participants were as follows: 193 male, 200 female, 13 non-disclosed. Their age ranged from 18 to 81 years (mean 45.2, SD 15.5): Silent Generation (75-92 years) = 3, Baby Boomers (56-74 years) = 130, Generation X (40-55 years) = 112, Generation Y / Millennials (24-39 years) = 119, Generation Z / Zoomers (<24 years) = 42. Concerning the level of education,  106 had a high school diploma or lower, 236 vocational training or a Bachelor's degree, and 64 were post-graduates. Several iterations with 16 pre-test participants served to enhance the comprehensibility of the questions and to reduce the duration to max. 30 minutes to avoid participants' fatigue. The survey was published and completed on Prolific on July 7, 2020. All participants were compensated with 3.75£, a price indicated as fair by Prolific\footnote{\url{https://researcher-help.prolific.co/hc/en-gb/articles/360009223533-What-is-your-pricing-}}.

\subsection{Ethical and Legal Considerations}
The study adheres to the University of Luxembourg's research ethics guidelines and the European Federation of Psychologists’ Associations’ code of ethics\footnote{\url{http://ethics.efpa.eu/metaand-model-code/meta-code/}}. In addition, the authorisation of the University's Ethics Review Board was obtained prior to the study. The survey gathered answers anonymously, and the questions did not inquire about information that would allow the identification of participants.

\subsection{Data analysis}
\subsubsection{Awareness of and concern about the influence of manipulative online designs (RQ1)} First, we calculated the mean, median and mode scores for the awareness ratings. We then computed a two-sided sign test to verify if the delta between the ratings referring to the participants (i.e., personal perspective) and their ratings referring to people in general (i.e., general perspective) was significant. Bivariate Pearson correlations were furthermore used to analyse how the personal awareness ratings correlate with the demographic data. The qualitative answers on awareness, harm, and worry were coded in MAXQDA\footnote{\url{https://www.maxqda.com/}} through an inductive approach. Researcher one coded 10 per cent (41 participants) of the sample and developed a set of codes. Researcher two coded the same set with the possibility to add codes. Non-agreement cases were discussed and codes adapted. The same procedure was repeated with another set of 41 participants. Since the inter-coder agreement reached 0.81 (Kappa Brennan \& Prediger), researcher one finalised the coding for the whole data set. The codes included \textit{use cases, influence objectives, influence types, harm types, concern types} and \textit{types of victims}. 

\subsubsection{Dark pattern detection (RQ2)}
The open answers to each example in the third part of the survey were coded through a deductive approach, by assigning a score depending on whether the participant identified the manipulative design element(s) correctly (no=0 / partly=0.5 / yes=1). Each example showed one main dark pattern, identified by the authors following the sources where the examples were obtained. Further plausible manipulative design elements found by the respondents were inductively included in the pool of correct or partially correct answers. Researchers one and two coded the answers of 10 per cent of the sample and developed the codebook in a shared document. Non-agreement cases were discussed and the codebook consequently adapted. The same procedure was repeated with another random set of 41 participants by researchers one and four. The inter-rater agreement reached a kappa of 0.77 (Kappa Brennan \& Prediger). Given the substantial level of agreement, researcher one finalised the coding for the whole data set. The quantitative data analysis was undertaken in Stata\footnote{\url{https://www.stata.com/}}(v.16.1). The dark pattern detection scores for each participant were summed (ranging from 0 to 9). Since it is not possible to draw a distinct line between high and low detection scores, we could not transform detection outcomes into binary variables. An OLS regression was hence chosen to control for significant differences deriving from age, educational level, use frequency of online services, and disposition to be influenced by online designs. 

\subsubsection{Likelihood to fall for dark patterns (RQ3)} Similar to the dark pattern detection scores, we summed the participants' ratings about their likelihood to be influenced by the proposed designs. We then ran an OLS regression to estimate the strength of association of participants' likelihood to be influenced with awareness (personal perspective), dark pattern detection, acceptability, and demographic data. Linear regression was again chosen because influence outcomes for the totality of the dark pattern examples could not be transformed into binary variables.\label{sec:methodology}

\section{Results}
The following section presents the results with regard to the three research questions introduced in Sec. \ref{subsec:research_q}, starting with the findings on people's awareness of the influence of manipulative designs online, followed by people's capacity to detect dark patterns. As a final step, both are examined as indicators for people's likelihood to be influenced by manipulative designs. For each part, differences associated with age, educational degree, use frequency of online services, as well as disposition to online manipulation, are addressed. 

\subsection{People's awareness of the influence of online designs on their choices and behaviour}
The first research question asked if people are aware of the influence of designs on their choices and behaviours in online services.

\paragraph{Awareness of influence} Regarding the ratings of the three question pairs in part one of the survey (Sec. \ref{meth-Part1:awareness}), the results reported in Table \ref{table:Awareness ratings} show that the participants were aware that online designs can influence their choices and behaviours (Me: mean = 1.05 SD 0.78, median = 1, mode = 1). The qualitative analysis of the answers reveals that the participants strongly associated influential online designs with well-known brands such as Amazon (mentioned by 103 participants), Netflix (40), Facebook (38), Instagram (22), eBay (17), Twitter (15), Youtube (13). They acknowledged that online designs may shape their spending behaviour (64 mentions), as well as their content consumption (45) and service choice (23), mainly through personalised contents and recommendations (78), as well as special offers (41). Some participants evoked social influence (22) as an effective strategy. Only a small number of participants cited specific design elements like visual appeal (24) or layout (17) as influencing factors. Some of them pointed to a website's or app's ease of use as a factor influencing whether they use it or not (43).

\paragraph{Awareness of potential harm} The participants were uncertain if manipulative designs online can cause them harm (Me: mean = 0.00 SD 1.10, median = 0, mode = -1), as shown in Table \ref{table:Awareness ratings}. The most frequently cited service category whose influence was considered harmful is social media (mentioned by 24 participants). The most prominent harm identified by participants was harm to themselves (135) both of psychological and physical nature. This was followed by mentions of financial harm (89), such as debt and unreasonable spending (51). Fewer participants evoked cybersecurity threats (31) or harm to their privacy (16). Some mentioned the dangers related to misleading information (28), and how these might influence people's opinions, values and attitudes (19) and cause damage to society (13).

\paragraph{Worries about manipulative designs}
As shown by the results in Table \ref{table:Awareness ratings}, the respondents were undecided and showed the tendency  not to worry about being manipulated by online designs (Me: -0.29 SD 1.07, median -1, mode -1): \textit{"I am not so personally worried about being manipulated, because I know myself well enough to question things and not get manipulated."(P78)}.
However, regardless of their own age, they evoked apprehension for vulnerable people (mentioned by 17 participants) and specifically for young people (14), the elderly (12), and children (11). For these people, they worried about the influence on spending behaviour (51), leading to financial losses (89). They furthermore expressed worry about the presence of false or misleading information (45), coupled with their understanding that online services only serve pre-filtered information (22) which influence people's opinions (41) and impede informed choices (20). Such ill-formed decisions might eventually cause harm to society (29) as well as to people's physical and mental health (23). There were also worries about cybersecurity threats (26). Finally, respondents found it worrisome that it is challenging to discern manipulative attempts (23), especially for vulnerable individuals.

\paragraph{Personal versus general perspective}
Several comments referring to concerns highlight that the participants were more worried for other people than themselves: \textit{“I consider myself very aware of these sort of things but someone else who has not a lot of internet experience or online shopping or believes whatever they see or are told will follow everything.”(P215)}. The results confirm this impression: the participants rated awareness, harm, and worry significantly higher when referring to people in general, as opposed to themselves (Table \ref{table:Awareness ratings}). 

\begin{table*}[htbp]
\begin{tabular}{p{1.1cm}p{4.1cm}p{4.1cm}p{4.1cm}}
& \textbf{1. Aware of influence} & \textbf{2. Aware of potential harm} & \textbf{3. Worried about}\\ 
 \hline
\textbf{Me} & 1.05 (SD 0.78)/1/1 & 0.00 (SD 1.10)/0/-1 & -0.29 (SD 1.07)/-1/-1\\
\textbf{People} & 1.30 (SD 0.59)/1/1 & 0.66 (SD 0.97)/1/1 & 0.60 (SD 1.01)/1/1\\
\hline
Sign test & (n=115, x>=98, p=0.5)=0.0000 & (n=183, x>=175, p=0.5)=0.0000 & (n=233,x>=222, p=0.5)=0.0000\\
\hline
\end{tabular}
\caption{Mean/median/mode scores of participants' rating of their 1) awareness of the influence and 2) the potential harm caused by manipulative designs online and 3) their degree of worry, with regard to themselves and people in general; values ranging from -2 \textit{strongly disagree} to 2 \textit{strongly agree}; the last row shows the p values for the two-sided sign tests between \textit{Me} and \textit{People} ratings.}
\label{table:Awareness ratings}
\vspace{-5mm}
\end{table*}

\begin{figure}[htbp]
    \centering
    \includegraphics[scale=0.077]{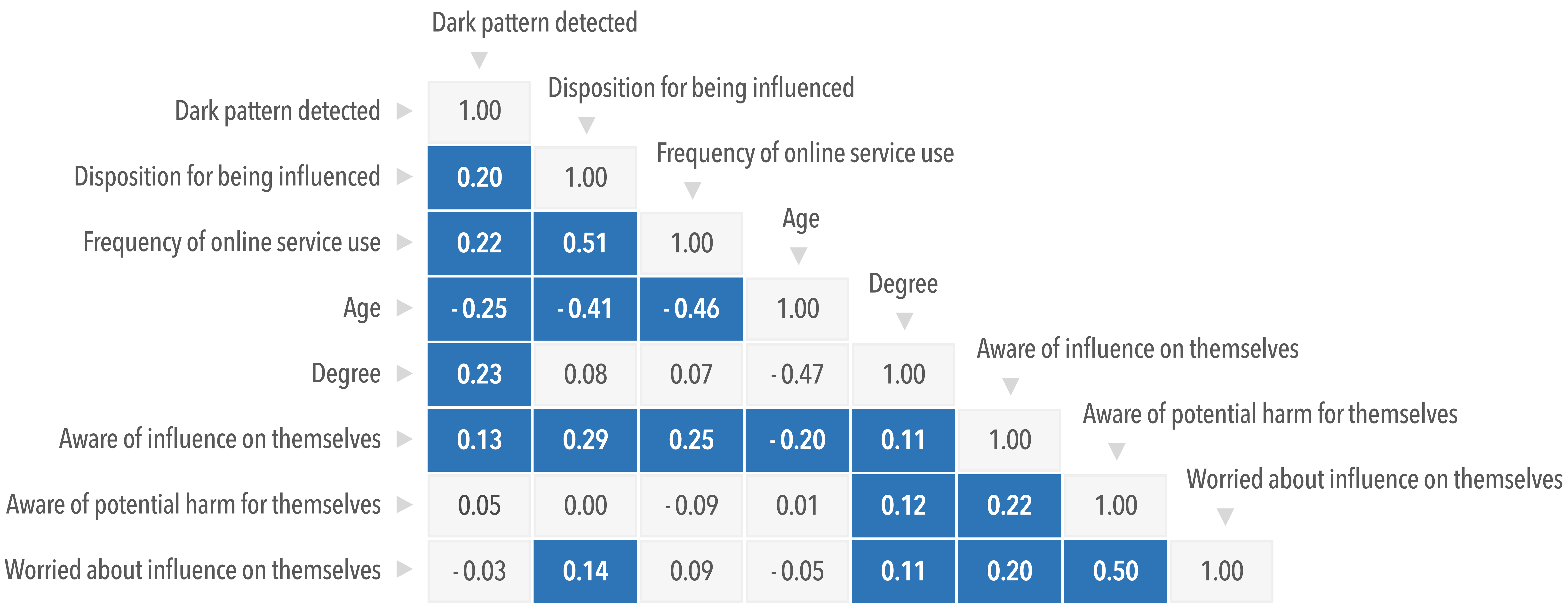}
    \caption{Correlation matrix for participants' awareness ratings and their individual characteristics, n=406, p < 0.05 with dark background.}
    \label{fig:awarenesscorrelations}
\end{figure}

\paragraph{Awareness by individual characteristics} The results displayed in Figure \ref{fig:awarenesscorrelations} show an inverse correlation between people's age and their awareness of online design's influence on themselves (r=-0.20, n=406, p=0.00). As some participants pointed out: \textit{“Being elderly I find it relatively easy to avoid being manipulated by these strategies. Technology has a place in my life but not an important place. I am not easily taken in.”(P250)}. Participants with higher education showed a slightly higher awareness of the influence of designs on their choices and behaviours (r=0.11, n=406, p=0.03), awareness of potential harm on themselves (r=0.12, n=406, p=0.01), as well as worry about the potential influence on themselves (r=0.11, n=406, p=0.02). Furthermore, the correlations indicate that those who use online services more frequently considered it more likely that manipulative designs influence their behaviour (r=0.25, n=406, p=0.00). Individuals with a higher disposition to be influenced also showed a higher awareness of their own likelihood of being influenced by manipulative designs (r=0.29, n=406, p=0.00). At the same time, they were also more worried about the influence on their choices and behaviour (r=0.14, n=406, p=0.01).

\paragraph{Summary of RQ1} On average, respondents are aware of online design's influence on their behaviour, especially on the type of online content they consume and the digital services they use. However, they are unsure if they can be harmed personally, even though they can name specific examples (e.g., frustration, anxiety, debt, loss of self-confidence). They are undecided on whether they should worry and are more concerned about other people than themselves. 

\subsection{People's ability to detect dark patterns}
\begin{figure}[htbp]
    \centering
    \includegraphics[scale=0.42]{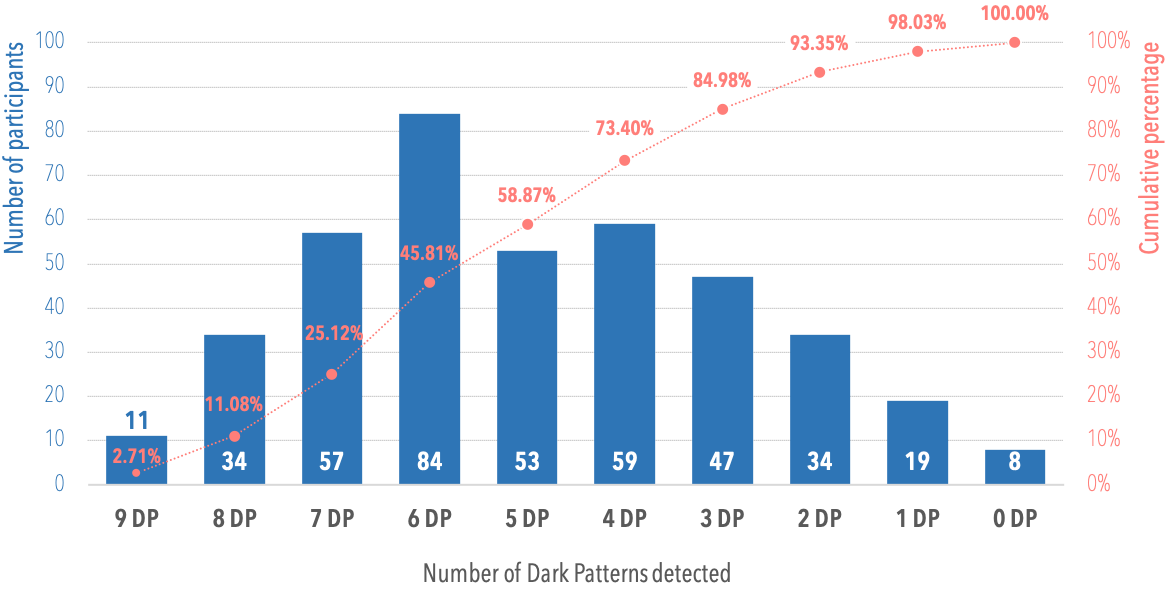}
    \caption{Frequency of dark pattern detection}
    \label{fig:frequencychart}
\end{figure}

Research question two sought to investigate whether individuals are able to recognise dark patterns. The results show that, when asked to look for elements that can influence users' choices and behaviour, 59\% of the participants were able to identify five or more dark patterns out of the nine interfaces correctly. One fourth recognised the dark patterns in seven, eight, or all nine interfaces (Figure \ref{fig:frequencychart}). As can be seen in Table \ref{table:DP data}, the interfaces including the dark pattern types \textit{trick question, pre-selection, loss-gain framing, hidden information and bundled+forced consent} were only recognised by half or less of the participants, while the majority of the participants correctly identified the interfaces containing a \textit{high-demand / limited time message and confirmshaming}.


\begin{table*}[htbp]
\begin{tabular}{p{5.4cm}p{0.8cm}p{0.8cm}p{0.8cm}r r}
& \multicolumn{3}{l}{Dark pattern detected} & Influential & Acceptable\\ 
Dark pattern name & \textit{no} & \textit{partly} & \textit{yes} & &\\
 \hline
a) Trick question / Pre-selection & \textbf{51}\% & 9\% & 40\% & -0.14(SD1.22)/0/1 & -0.73(SD1.12)/-1/-1\\
b) Loss-gain framing / Confirmshaming & 27\% & 19\% & \textbf{53}\% & -0.03(SD1.26)/0/1 & -0.79(SD1.06)/-1/-1\\
c) Pre-selection / Loss-gain framing & \textbf{51}\% & 11\% & 38\% & 0.00(SD1.20)/0/1 & -0.50(SD1.08)/-1/-1\\
d) Hidden information / Trick question & \textbf{64}\% & 20\% & 16\% & 0.36(SD1.21)/1/1 & -1.17(SD1.06)/-2/-2\\
e) Bundled / Forced consent & \textbf{50}\% & 36\% & 14\% & 0.36(SD1.15)/1/1 & -0.64(SD1.06)/-1/-1\\
f) High-demand / Limited-time message & 5\% & 11\% & \textbf{84}\% & 0.39(SD1.25)/1/1 & -0.39(SD1.11)/0/-1\\
g) Confirmshaming & 27\% & 2\% & \textbf{71}\% & -0.85(SD1.15)/-1/-2 & -0.22(SD1.10)/0/1\\
h) Hidden information / False hierarchy & 42\% & 4\% & \textbf{54}\% & -0.49(SD1.28)/-1/-1 & -1.00(SD1.09)/-1/-2\\
i) Auto-play & 40\% & 11\% & \textbf{49}\% & 0.72(SD1.05)/1/1 & 0.74(SD0.87)/1/1 \\
 \hline
\end{tabular}
\caption{Dark pattern detection percentages for the 9 dark pattern interfaces in the survey and mean/median/mode scores of participants' evaluation of their likeliness to be influenced by the dark pattern and the acceptability of the strategy (from -2 = \textit{strongly disagree} to 2 = \textit{strongly agree}).}
\label{table:DP data}
\vspace{-5mm}
\end{table*}

\begin{figure}[htbp]
    \centering
    \includegraphics[scale=0.45]{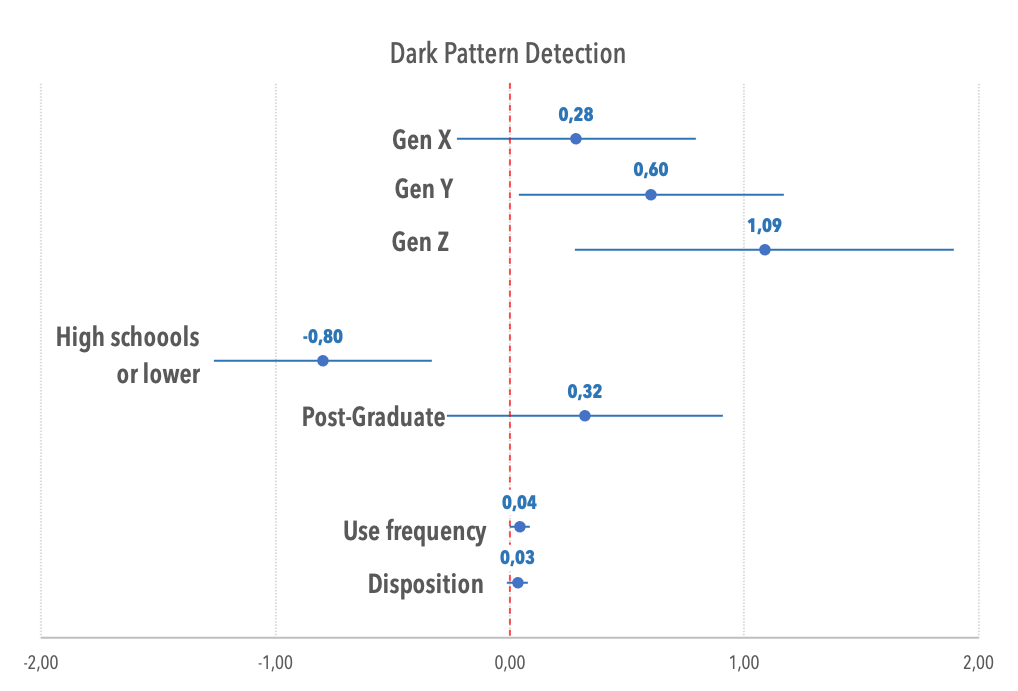}
    \caption{Plot from linear regression with robust standard errors for dark pattern detection; n=406, Adj $R^2$=0.097, BIC=1754.781.}
    \label{fig:detection_plots}
\vspace{-3mm}    
\end{figure}

\paragraph{Dark pattern detection and individual characteristics} Using OLS regression analysis to model the inter-relationship between the detection of dark patterns and associated factors (Figure \ref{fig:detection_plots}), it emerges that younger people could identify a higher number of dark patterns than the older Baby Boomer+ generation\footnote{Due to the low number of Silent Generation participants in the survey, the reference category combines the Baby Boomer generation (people born between 1946-1964) and the older “Silent Generation” (people born between 1928-1945).} net of education, use frequency, and disposition: Millenials/Gen Y: coef. .60 (95\%CI: 0.04 to 1.16); Zoomers/Gen Z coef. 1.09 (95\%CI: 0.35 to 1.83). Generation X is not better or worse than the older Baby Boomer+ generation, as indicated by the regression results coef. 0.28 (95\%CI: -0.25 to 0.81). Regarding education, participants with a high school degree or lower detected less dark patterns (coef. -0.80 (95\%CI: -1.26 to -0.33)), compared to participants with a Bachelor's degree or vocational training. However, participants with higher degrees than Bachelor were neither better nor worse at identifying manipulative design strategies (coef. 0.32 (95\%CI: -0.24 to 0.88)). This suggests that the Bachelor/vocational training level is a threshold below which recognition rates are lower. The regression analysis also indicates a slight positive correlation between online use frequency and the number of dark patterns detected (coef. 0.04 (95\%CI: 0.00 to 0.09)), but no significant correlation for disposition to manipulation and dark pattern detection (coef. 0.03 (95\%CI: -0.01 to 0.09)).

\paragraph{Dark pattern detection and awareness}
The correlation matrix (Figure \ref{fig:awarenesscorrelations}) shows a positive correlation between recognition of dark patterns and awareness of manipulative designs' influence (r=0.13, n=406, p=0.01). However, many participants were surprised that they were unable to recognise certain manipulative designs:

\textit{"I like to think I am pretty 'switched on' when it comes to avoiding being manipulated online and this highlighted to me how much I sign away at the click of the button!"(P150)}

\textit{"There were a few elements that I missed, but were obvious when pointed out. It shows how easy it is to be manipulated, even when one thinks they are aware."(P358)}

\paragraph{Summary of RQ2} We conclude that people are able to recognise dark patterns, but there is variation across dark pattern types. Younger age (<40 y.), as well as education levels above high-school degree, are positively correlated with this ability.  

\subsection{People's likelihood to be influenced by dark patterns}
\begin{figure}[htbp]
    \centering
    \includegraphics[scale=0.22]{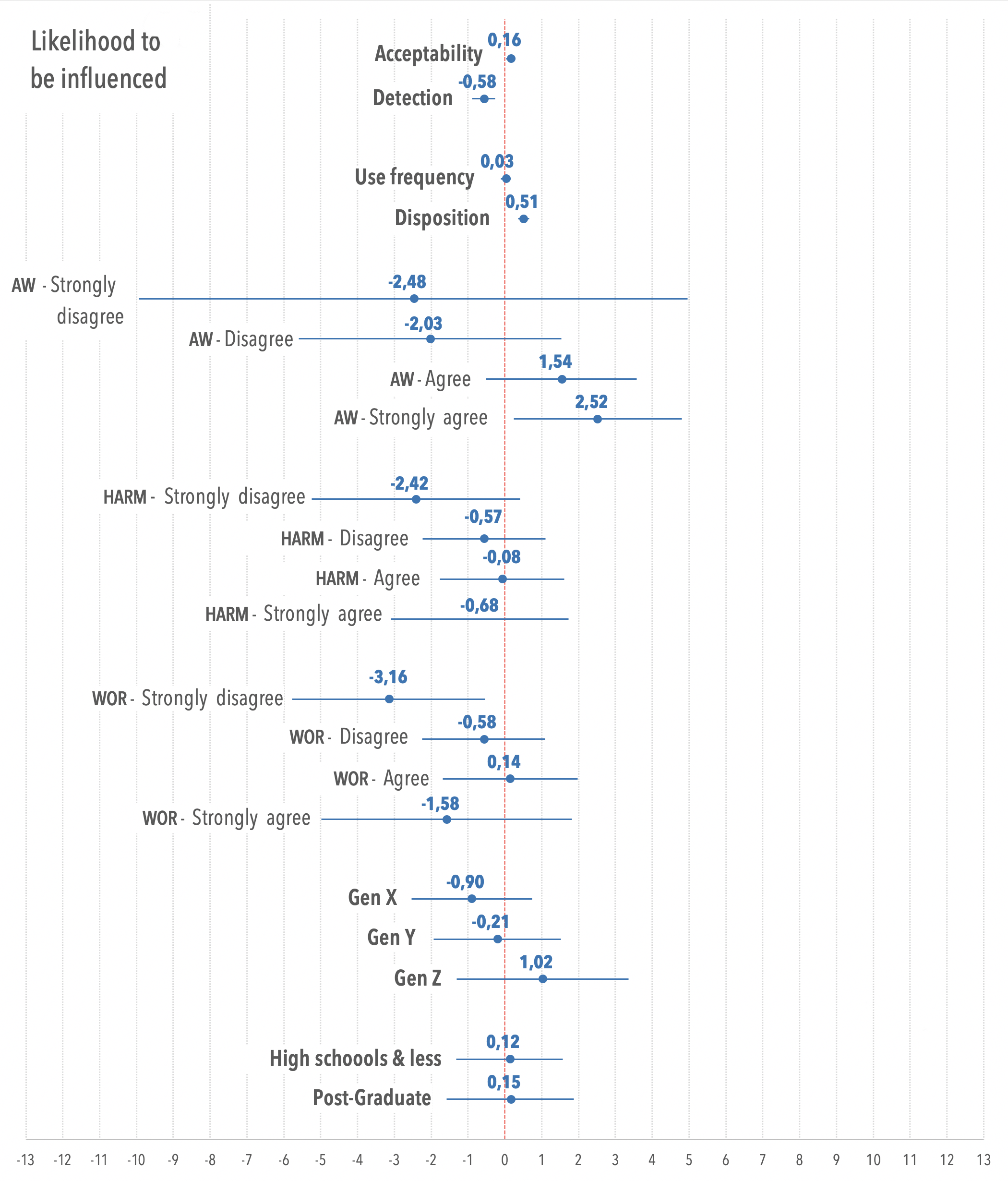}
    \caption{Plot from linear regression with robust standard errors for likelihood for being influenced by dark patterns; n=406, Adj $R^2$=0.3039, BIC=2723.14; AW = awareness, WOR = worry.}
    \label{fig:likelihood_plots}
\vspace{-3mm}    
\end{figure}

The third research question sought to investigate if higher awareness, as well as a higher capability to detect manipulative designs, make people less likely to be influenced. We ran an OLS regression analysis to model the inter-relationship between the participants' self-reported influence-likelihood rating and the associated factors (Figure \ref{fig:likelihood_plots}).

\paragraph{Likelihood of being influenced and the capacity of dark pattern detection} The data shows a slight inverse correlation between participants' dark pattern detection capability and their influence-likelihood rating (coef. -0.58 (95\%CI: -0.90 to -0.26)). This indicates that people who recognise manipulative designs more easily consider themselves slightly less likely to be influenced by them.

\paragraph{Likelihood of being influenced and dark pattern acceptability} In all tested designs, participants were on average uncertain whether the dark patterns would influence their behaviour and whether they find them acceptable (Table \ref{table:DP data}). Those who considered these designs more acceptable also reported being slightly more influenced in their behaviour (coef. 0.16 (95\%CI: 0.04 to 0.28)). Interestingly, a design's admissibility was not necessarily related to its influence strength. For example, about half of the participants recognised both \textit{(h) hidden information / false hierarchy} and \textit{(i) auto-play}. After receiving an explanation about what could be considered manipulative in both interfaces, the participants tended to find \textit{(i) auto-play} more influential than \textit{(h) hidden information / false hierarchy}. However, they deemed \textit{(i) auto-play} more acceptable than \textit{(h) hidden information / false hierarchy}. 

\paragraph{Likelihood of being influenced and awareness} Participant comments at the closure of the survey reflect that awareness is not a significant predictor for participants' likelihood to be influenced by manipulative designs.

\textit{"I think I’m aware of most manipulative practices but there are certain applications like video streaming and booking accommodation where I am definitely manipulated, even when I am aware of it. It’s ridiculous!"(P139)}

\textit{"I feel I am quite aware of some of the subtleties of advertising and suggestion but there were elements I hadn't even considered may be unconsciously influencing my choices."(P222)}

Indeed, only respondents who strongly believed that online designs can influence them, also deemed it likely to be influenced by the designs in the tested interfaces (coef. 2.52 (95\%CI: 0.24 to 4.81)). Conversely, people who strongly disagreed with being worried about the influence of online designs on themselves considered it also unlikely to fall for the designs in the tested interfaces (coef. -3.16 (95\%CI: -5.77 to -0.54)).

\paragraph{Likelihood of being influenced and individual characteristics}
The regression analysis shows no significant correlation between age, education, or online use frequency and the self-reported likelihood to be influenced by manipulative designs. However, there is a positive correlation between participants' disposition to online manipulation and their ratings concerning the influence likelihood (coef. 0.51 (95\%CI: 0.36 to 0.66)), which hints at a certain degree of inability to resist manipulative designs.

\paragraph{Summary RQ3} We conclude that people who recognise manipulative designs with more ease report, on average, a lower likelihood of being influenced by them. However, whether people are very aware of online manipulative attempts or not makes, on average, no difference in terms of their likelihood to be influenced by such designs.\label{sec:results}

\section{Discussion}
\begin{figure*}
    \centering
    \includegraphics[scale=0.6]{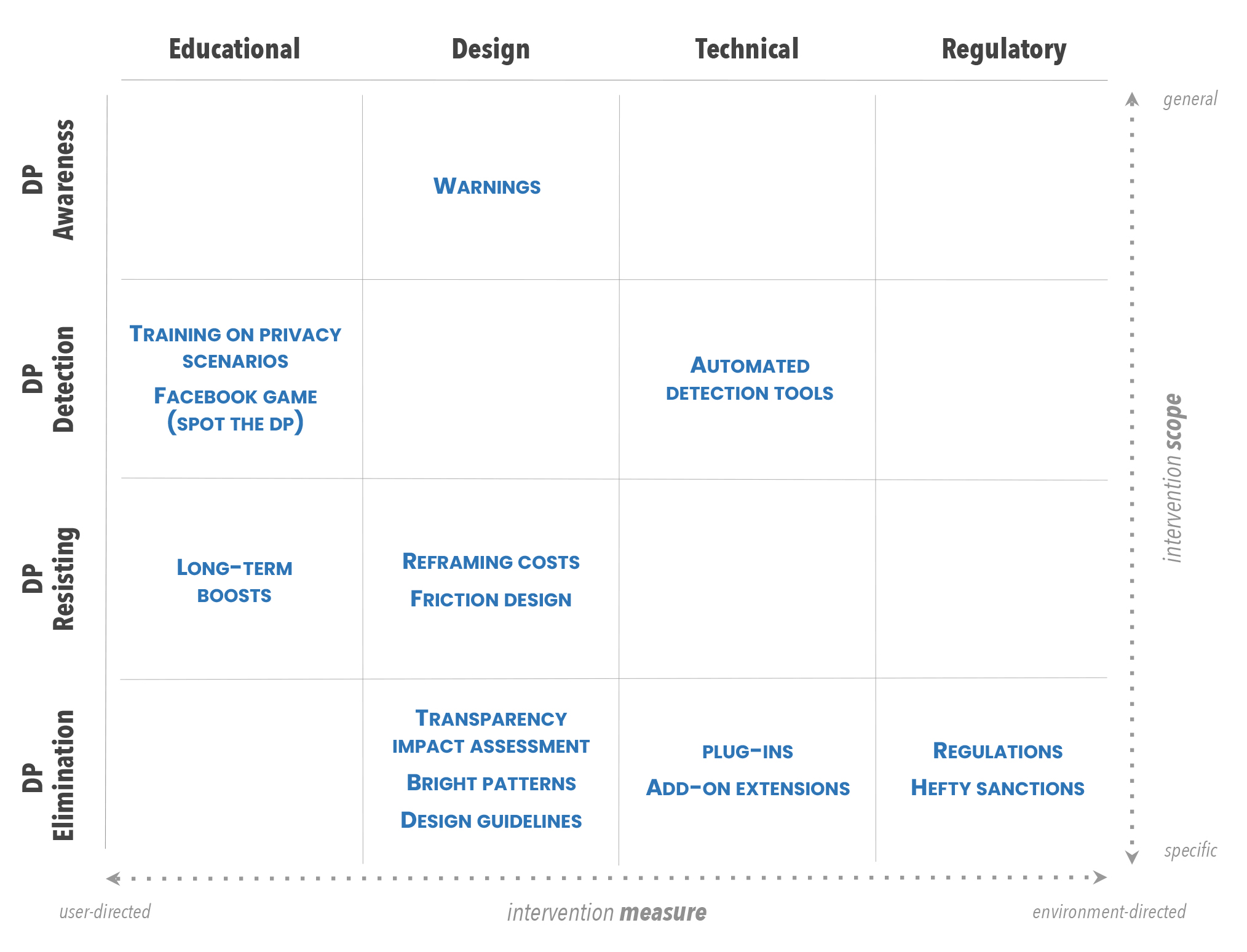}
    \caption{Intervention spaces for counteracting dark patterns.}
    \label{fig:interventionspaces}
\end{figure*}

We discuss the results in light of the interventions that could be put in place to counteract dark patterns. Interventions can aim to (i.e., intervention scope) a) raise awareness of the existence and the risks of dark patterns, b) facilitate their detection, c) bolster the resistance towards them, or d) eliminate them from online services. Interventions can act on the user or the environment (i.e., intervention measures): educational interventions favour users' agency, regulatory interventions tend to protect the user, technical and design interventions are situated in-between. This distinction can serve to identify the actors (e.g., design practitioners, researchers, educators, regulators) that should implement the interventions and devise appropriate evaluation indicators. The resulting matrix is shown in Figure \ref{fig:interventionspaces}. 

\subsection{Raising awareness}
The results indicate that people are generally cognizant that digital services can exert a detrimental influence on their users, but fail to understand how manipulative designs can concretely harm them. Individuals' lack of sufficient concern does not impact their ability to spot dark patterns. However, it may impact their motivation to counter them, like taking a few extra steps to select the less privacy-invasive option in consent dialogues. Moreover, people are more worried about the danger represented by dark patterns for other people than for themselves, thus confirming previous assumptions \cite{west2008psychology}. Warnings \cite{grassl2020dark} are a design intervention that can make threats salient and concrete (e.g., about financial losses following mindless purchasing decisions) and counterbalance the tendency to underestimate online threats due to hyperbolic discounting and optimism bias. However, warnings become rapidly ineffective as users get habituated (i.e., warning fatigue \cite{Akhawe2013alice}) and need to mutate continuously to continue capturing users' attention \cite{Bravo2014harder}. 

\subsection{Facilitating detection}
The study results on dark pattern detection show significant variations across the proposed designs. A majority of users recognised confirmshaming and the high-demand / limited-time message (similar to \cite{Shaw2019wise_nudge}), whilst dark patterns based on deception strategies (e.g., trick question, loss-gain framing and hidden information), together with the pre-selection nudge and forced consent, were scarcely recognised. Although such findings only concern a specific implementation of the dark pattern and cannot be generalised to the category, they may suggest that certain dark patterns are intrinsically more difficult to spot. For instance, the omission of information is a shrewd deceptive strategy. It requires users to have a correct mental model of expectations, coupled with high cognitive activation \cite{boush2009deception}, to notice the absence of certain elements. In our facial recognition example which was based on loss-gain framing, many respondents were simply unaware of the unbalance in the presentation of the arguments (\textit{"I don't think this one is manipulative, it's just explaining the benefits of using face recognition, and the possible drawbacks of not using it"(P169)}) and of the entailed risks (\textit{"Turning on Facial Recognition, is it good or bad??"(P141)}). This probably explains why this dark pattern was rarely identified. When respondents mentioned possible risks, their answers revealed wrong mental models about the drawbacks of facial recognition, like installation of malware, target advertisement, or unlawful surveillance. In such cases, educational measures such as training on cause-and-effect data privacy scenarios can act successfully to sharpen manipulation detection abilities. As for what concerns the poor detection of pre-selection and forced consent, a plausible explanation is that users have grown accustomed to such designs. However, a dedicated study could demonstrate which design attributes \cite{Mathur2021what} make certain dark patterns harder to spot.

It would also be useful to investigate folk models about dark patterns: mental models that are not necessarily accurate in the real world and lead to erroneous decision-making \cite{wash2010folkmodel}. Complementarily, it should be further researched which attributes trigger users' scepticism  towards interfaces and activate a more elaborate mode of thought (i.e., counterfactual thinking \cite{boush2009deception}) that disposes them to recognise potential manipulation attempts. For instance, respondents in \citet{Bhoot2020IndiaCHI} indicated sudden interruptions and excessive ads as elements activating scepticism. However, such research also found that users hardly notice a manipulation attempt when the interface is appealing. This result is in line with previous work, demonstrating that people base their online trust judgements on cues, such as the visual ones \cite{Wang2004trustworthiness}. On this note, what has been learnt in anti-phishing research about the cues that evoke distrust in professional-looking e-mails can be of use to determine how to activate the "critical persuasion insight" \cite[p. 114]{boush2009deception}. Many comments of our respondents hinted that activities like "spot the dark pattern" can serve as an eye-opener: \textit{"This has been a great survey and it has certainly made me more aware of certain things that I have not noticed in the past. I will be keeping an eye out for such things going forward" (P214)}. Similar gamified experiences\footnote{See e.g.  \url{https://cookieconsentspeed.run/}, a game where users need to navigate ambiguous options and distrust obvious buttons in cookie dialogues.} integrated into major digital services (e.g., a Facebook game) could strengthen the motivation to learn how to spot dark patterns in real settings, without the cognitive cost of transferring skills learnt in training to the context of digital services.

Concerning technical interventions, algorithms and applications that  automatically identify, flag, and even classify potentially illegal practices at large scale should be developed on the model of \cite{mathur2019dark,nouwens2020dark,lippi2019claudette}, to expedite watchdogs' supervising tasks and provide proof to consumer advocates. Such tools need a large pool of reliable data to carry out the recognition and categorisation of manipulative attempts that are challenging even for humans. To this end, we are currently assembling a corpus of dark pattern interfaces published on Reddit\footnote{https://www.reddit.com/r/assholedesign/ and https://www.reddit.com/r/darkpatterns/} and Twitter\footnote{The tweets containing hashtags like \#darkpattern} by social media users.

\subsection{Bolstering resistance}
Even though there are no significant correlations in milder ratings, those respondents who declared to be very likely influenced also showed a higher awareness of this possibility and greater concern. This hints that awareness of one's own vulnerability does not automatically trigger better self-defence against manipulative influences. Design interventions can enhance users' appraisal of the effort it takes to cope with certain dark patterns -- for example, indicating the time it takes to unsubscribe when it is an overly complex procedure (see e.g., Amazon Prime \cite{Forbruker2021amazon}). Reframing the costs of falling prey to dark patterns in personally relevant terms, as proposed by \citet{Moser2019impulse} to contrast impulse buying, may also be considered -- for instance, by converting the time spent on infinite scrolling into other pleasurable activities. User research can determine what is valuable for users, as it emerges from our respondents (e.g., P278: \textit{"It [is] such a waste of precious time, that can be used in reading, personal time, ecercise [sic] some more benefitial [sic] for the individual"}.) Friction designs can disrupt automatic behaviour with positive effects by introducing small obstacles that create a brief moment of reflection and stimulate more mindful interactions \cite{Cox2016frictions}. Already in use to induce more secure online behaviour \cite{Distler2020friction} and proposed to counter irrational spending behaviour \cite{Moser2019impulse}, friction designs are now widespread on streaming services (e.g., YouTube, Netflix) to counter binge-watching. Similar nudges could oppose infinite scrolling, defaults, and mindless consent to data sharing and extensive online tracking. 

The study participants who recognised more dark patterns also reported a lower likelihood of being influenced by them. This suggests that the ability to recognise a threat is intimately related to the ability to protect oneself \cite{Rogers1997protection}. The disclaimer that manipulative elements may be contained in the interfaces had the effect of activating participants' counterfactual thinking and encouraged a more reflective way of processing information. An educational intervention like long-term boosts can build manipulation-protection abilities and empower people to apply them without having to resort to deliberate thinking  in the long-run \cite{Hertwig2017boosting}, like the procedural rules proposed in \citet{grassl2020dark}: "every time I encounter a cookie consent dialogue, I search for the 'refuse all' button". 

However, the cost of resisting dark patterns varies depending on whether they employ coercive, nudging, or deceptive strategies, and on their specific implementations. Coercive dark patterns (e.g., forced consent) are inescapable: if someone desires to use a service that integrates a coercive design, they do not have the possibility of avoiding it (i.e., `take-it-or-leave-it'). Dark patterns preventing individuals from accomplishing a task are similarly daunting, since opposing them would come with a high cognitive cost. For instance, only a (motivated) minority of website visitors is willing to take additional steps to adjust their preferences on cookie dialogues \cite{utz2019informed}. Nudging strategies may be implemented variously and thereby exert more or less influence on users (e.g., mild vs aggressive dark patterns \cite{luguri2019shining}). Deceptive strategies, on the other hand, can be resisted only through the activation of counterfactual thinking.

\subsection{Eliminating dark patterns from digital services}
Dark patterns are present in more than 10\% of global shopping websites \cite{mathur2019dark}, in almost 90\% of cookie consent dialogues of the top 10000 websites in the UK \cite{nouwens2020dark} and more than 95\% of the 200 most popular apps \cite{di2020ui}. To respond to such pervasiveness, technical solutions that ease autonomous decision making can be devised, like Do Not Track\footnote{\url{https://allaboutdnt.com/}}, the add-on extension Consent-O-Matic\footnote{\url{https://addons.mozilla.org/en-US/firefox/addon/consent-o-matic/}.} or browser plug-ins that disable other plug-ins, e.g., those that create scarcity messages. 

Digital nudges that counteract dark patterns (i.e., bright patterns) by, for example, making the privacy-savvy option more salient, modify the environment where users make choices. There is a rich literature concerning design nudges that enhance privacy decision making \cite{acquisti2017nudges}, including personalised nudges adapted to individual decision making styles \cite{Peer2020nudgeright}. \citet{grassl2020dark} found that bright patterns nourish users' perception of lack of control, though, as they act on unreflective behaviour in the same way as dark patterns. Coercive and deceptive dark patterns (e.g., forced consent, trick questions), though, cannot be defeated through digital nudges. A complementary design intervention consists of promoting good practices through the publication of design guidelines \cite{acm2020guidelines,age_designcode} and the involvement of companies in problem-solving activities on concrete case studies \cite{CNIL_donnees}. Ethical design tool-kits\footnote{E.g., \url{https://ethicalos.org}.} can be employed to foresee the consequences of certain designs comprehensively. At the same time, persuasive technology heuristics (e.g., \cite{Kientz2010heuristic}) may be adapted to assess the potential manipulative effects of digital products even before their release. Building on such initiatives and \cite{Meske2020ethical,Mathur2021what}, we plan to develop a standardised transparency impact assessment process for interface design. 

However, given the omnipresence of dark patterns on online services, it is somewhat unrealistic to expect businesses to implement such interventions on their own: economic incentives and regulatory interventions should complement other proposed actions. Legal safeguards should apply more stringently, as many dark patterns are unlawful in the EU under consumer law (e.g., omission of information, obstruction to subscription cancellation) \cite{acm2020guidelines,Forbruker2021amazon,Leiser2020regulatory} and the data protection regime (e.g., forced consent, loss-gain framing) \cite{EDPB_consent2020,forbrukerradet_deceived_2018}. Stiff penalties can furthermore act as a deterrent: in France, for instance, Google has received fines for a total of 150 million € due to invalid consent design and elicitation \cite{CNIL2019google,CNIL2020google}. Empirical research demonstrating the presence, diffusion and effects of manipulative designs might have an impact on legal enforcement: cookie consent dialogues increasingly offer privacy-by-default options as a result of case law (e.g., the landmark case Planet49 \cite{planet49}) and, conceivably, of intense academic scrutiny. The threat of more stringent regulations (e.g., the US Social Media Addiction Reduction Technology Act) and public pressure (e.g., derived from the popularity of documentaries like ``The social dilemma'') may even encourage self-regulation.

\subsection{Targeting interventions - older vs younger generations}
Our results highlight that older generations are not only less able to recognise manipulative attempts, but they are also less aware that their choices and behaviour can be influenced. This could be problematic, as perceived vulnerability to harm is a key factor to trigger self-protection \cite{Rogers1997protection}. The combination of lack of awareness and lack of capability makes dark patterns' effects particularly dangerous for older adults, as they struggle to adapt their learned self-protection abilities to evolving (digital) environments \cite{boush2009deception}, echoing findings about online misinformation \cite{Brashier2020aging}. That said, it is arguably easier to define \textit{ad hoc} protections addressed to younger populations (e.g., like the ICO's ``Age Appropriate Design'' code of practice \cite{age_designcode}) than to older ones: how would targeted safeguards for over 40s be enacted and received? Moreover, our findings do not suggest any significant correlation with the likelihood to be influenced, although it would be worthwhile to expand research in this direction. The study results show that an age lower than 40 years and an education level higher than high school diploma constitute a critical threshold for recognising dark patterns and could indicate that the other part of the population is sensibly less likely to be aware of manipulative attempts online.\label{sec:discussion}

\section{Limitations}
The choice of a large representative sample of the UK population for this survey has the objective of generalizing the findings to the whole population. However, given that Prolific is an online platform, participants are probably more accustomed to online designs than the average (e.g., the distribution of the participants' online use frequency shows a positive skew). Therefore, our results might overestimate what a less tech-savvy UK population is aware of when it comes to manipulative designs online. It would also be interesting to find out whether the study would achieve different conclusions in other countries. It should be further mentioned that the participants' likelihood to be influenced by manipulative designs was derived from a self-reported measure and does not necessarily reflect actual behaviour, which makes the measure an approximation and invites further research.

Following the ratings of awareness, harm, worry in part one of the survey, the participants were invited to cite examples. Whilst we explicitly asked about the influence of manipulative designs, people also cited manipulative content (e.g., fake news), signalling that it is not obvious for people to distinguish form from content. We thus assume that the participants' awareness of and concern about the influence caused by manipulative designs may be lower than indicated by the results. Concerning the dark pattern detection activity, we are aware that explicitly searching manipulative design elements does not entirely correspond to a real use situation. We sought to counterbalance the effect through the time limit and the allusion that some interfaces would not contain manipulative elements. That said, we also estimate that such settings could correspond to a real-world scenario where people remark something odd and activate counterfactual thinking \cite{boush2009deception}.\label{sec:limitations}

\section{Conclusions}
Manipulative designs are a growing threat in the online environment. Practitioners and researchers from multiple domains (HCI, computer science, law, etc.) currently seek to expose and counteract their influence on user behaviour. Yet, to shield users effectively, it is essential to understand their capabilities when confronted with manipulative designs. This study shows that individuals are aware of manipulative designs' potential influence on their behaviour and rather capable of recognising such designs. While an inverse correlation between dark pattern design recognition and participants’ likelihood to be influenced was found, the level of awareness did not play a significant role in predicting their ability to resist manipulative designs. This finding implies that raising awareness on the issue is not sufficient to shield users from the influence of dark patterns.

Our discussion presented a palette of interventions (i.e., design, technical, educational, and regulatory measures) meant to heighten people's awareness of manipulative design practices, ease their detection, strengthen people's resistance to them, or root them out. We believe that design measures (like frictions and bright patterns) and technical solutions (like automated dark pattern detection applications) should be further investigated, together with assessment tools, economic incentives and regulatory solutions.

To complement scholars' and authorities' views on the issue, we suggest exploring established dark pattern attributes in combination with the user perspective as part of future work. Only by understanding which (combinations of) attributes are commonly perceived as unrecognisable, irresistible and/or unacceptable by the users, we can devise appropriate interventions. Moreover, the exploration of user perceptions can help establish what is deemed legitimate and what is not by end-users without taking a normative stance. Looking at dark patterns from the user perspective shows that they are a problem with many variables. As such, they require that a variety of actors teams up to devise a kaleidoscope of interventions. Designers should be on the front line to help tame the monster they contributed creating.\label{sec:conclusions}

\begin{acks}
This publication is a first step of the project Decepticon (grant no. IS/14717072) supported by the Luxembourg National Research Fund (FNR). We would like to thank the anonymous reviewers of DIS 2021 and CHI 2021 for their helpful comments and all those who have helped us refine the study design.
\end{acks}

\bibliographystyle{ACM-Reference-Format}
\bibliography{bibliography}

\end{document}